# Sustainability and Fairness Simulations Based on Decision-Making Model of Utility Function and Norm Function

Takeshi Kato[1], Yasuyuki Kudo[1], Junichi Miyakoshi[1], Jun Otsuka[2], Hayato Saigo[3], Kaori Karasawa[4], Hiroyuki Yamaguchi[5], Yoshinori Hiroi[6] & Yasuo Deguchi[2]

[1] Hitachi Kyoto University Laboratory, Open Innovation Institute, Kyoto University, Kyoto, Japan

[2] Department of Philosophy, Graduate School of Letters, Kyoto University, Kyoto, Japan

[3] Faculty of Bioscience, Nagahama Institute of Bio-Science and Technology, Shiga, Japan

[4] Department of Social Psychology, Graduate School of Humanities and Sociology, The University of Tokyo, Tokyo, Japan

[5] Department of Behavioral and Health Sciences, Graduate School of Human-Environment Studies, Kyushu University, Fukuoka, Japan

[6] Kokoro Research Center, Kyoto University, Kyoto, Japan

Correspondence: Takeshi Kato, Hitachi Kyoto University Laboratory, Open Innovation Institute, Kyoto University, Kyoto 606-8501, Japan.

**Abstract**

We introduced a decision-making model based on value functions that included individualistic utility function and socio-constructivistic norm function and proposed a norm-fostering process that recursively updates norm function through mutual recognition between the self and others. As an example, we looked at the resource-sharing problem typical of economic activities and assumed the distribution of individual actions to define the (1) norm function fostered through mutual comparison of value/action ratio based on the equity theory (progressive tax-like), (2) norm function proportional to resource utilization (proportional tax-like) and (3) fixed norm function independent of resource utilization (fixed tax-like). By carrying out numerical simulation, we showed that the progressive tax-like norm function (i) does not increase disparity for the distribution of the actions, unlike the other norm functions, and (ii) has high resource productivity and low Gini coefficient. Therefore the progressive tax-like norm function has the highest sustainability and fairness.

## 1. Introduction

In various countries of the world, various problems e.g., social problems, disparity and inequality, energy problems related to dependence on fossil fuels and nuclear power accidents, and environmental problems related to global warming and pollution, are emerging. To solve these problems and aim for a sustainable, fair and inclusive society, social norms in a broad sense such as fairness, equity, goodness, justice, obligation, morality and ethics ought to be considered. The focus of our research was on how to incorporate social norms and ethics into rational decision-making in economics and political science, how to understand and diagnose actions of individuals and groups in the real world, and how to intervene to make social systems better.

The major challenge faced in using a practical approach to solving social problems is that, one ought to both interprete and diagnose individual and group actions in the real world from the perspectives of both aspects of utility and norm, i.e. is writing a prescription while making a prognosis, and determining how to intervene into the social system, i.e., how to, so to speak, medically treat the social system in the clinical field. Conventional social intelligence paradigms are based on physical models that explain phenomena and predict the future based on analysis of data obtained from the phenomena, and also based on historical models that explain the origins of disasters and the prevention future disasters. However, human economy and society are complex systems composed entirely of various interacting components. They are also considered as autopoietic systems that bring about a cyclic network between components and generational changes in the components, as described by Niklas Luhmann in his social systems theory (Luhmann, 1996). Dealing with human society, which is a complex and constantly changing cyclic system, requires clinical medical models that carry out interventions based on continuous diagnosis and prognosis, rather than reductionist physical models or historical models that rely on transient phenomena, and necessitates the building of co-evolutionary relationships between the real world and practical intervention. Therefore, prior to making a diagnosis and handing out prescriptions, we sought to clarify what social norms were needed by a sustainable and fair society, and, in terms of economics, what social norms ought to be set against general utility theories.

In "The Wealth of Nations", Adam Smith said, "He intends only his own gain, and he is in this, as in many other cases, led by an invisible hand to promote an end which was no part of his intention. Nor is it always the worse for the society that it was not part of it. By pursuing his own interest, he frequently promotes that of the society more effectually than when he really intends to promote it." (Smith, 1776). In support of this prediction, Kenneth Arrow and Gérard Debreu et al. proved that society reaches its optimum state when the contract is complete, and all individuals pursue their own self-interests as the first fundamental theorems of welfare economics.

Conversely, in "The Theory of Moral Sentiments", Smith stated, "How selfish soever man may be supposed, there are evidently some principles in his nature, which interest him in the fortunes of others, and render their happiness necessary to him, though he derives nothing from it, except the pleasure of seeing it." (Smith, 1759). Regarding how a complete contract cannot exist in real society, Arrow said "I want, however, to conclude by calling attention to a less visible form of social action: norms of social behavior, including ethical and moral codes. I suggest as one possible interpretation that they are reactions of society to compensate for market failures. " (Arrow, 1970).

To illustrate examples of previous research on social norms that have overtaken the discussions of Smith and Arrow: philosopher Joseph Heath proposed the adoption of normative appropriateness as deontic constraints into a rational choice model based on the expected utility theory for decision making (Heath, 2008), Economist Kaushik Basu acknowledged the existence of subjective moral costs in decision making (Basu, 2010), Economist Masahiko Aoki showed that community norms emerge intrinsically when the cost for cooperation is smaller than the loss from social ostracization, by linking the commons game and social exchange game (Aoki, 2001), and Economist Samuel Bowles forwarded the importance of moral motivation and social preference in markets based on incomplete contracts (Bowles, 2017). Their common viewpoint is that they all point out the need to incorporate social norms, which include morals and ethics, not only into economic utility, but also into individual decision-making models.

Previous research utilized different approaches. The expected utility theory, prospect theory, amongst other theories, as individual decision-making models in microeconomics; The expected utility theory discussed risk-averse type and risk-seeking type utility function forms in making decisions to maximize the expected value of utility for choices under conditions of uncertainty, whereas, the prospect theory explained reference points and loss aversion (non-linearity) in making decisions based on utility functions that are assigned with weights for probability weighting function (e.g. Gilboa, 2010). The concepts of altruism and reciprocity through various game experiments in experimental economics; The game theory expressed utility functions, for example, using a prisoner's dilemma payoff matrix, ultimatum game logic tree, and public goods game allocation rules, and discusses the Pareto efficiency and competitive equilibrium for decision-making strategies (e.g. Bowles & Gintis, 2013). The concepts of axiomatic characterization of institutionalized mechanisms as normative and ethical approaches in welfare economics; Welfare economics proposed theorems pertaining to competitive equilibrium and Pareto efficiency based on completeness and transitivity axioms, and also formulated mappings from sets of individuals and economic environments to sets of individual goods and capabilities using ordinal utility functions based on profiles of individual preference orders and resource-use capabilities, to come up with standards pertaining to game forms and social welfare as institutionalized mechanisms (e.g. Sen, 2017). These theories have contributed significantly to society and economics. This study, however, aims to determine the concrete effects of explicitly incorporating cardinal norm functions in decision-making models, towards creating a better society (i.e. to sustainability, fairness, disparity, etc.), and to identify how to foster norms in social systems when carrying out social practices.

In this paper, in Chapter 2 we introduce a value function that includes utility function of methodological individualism and norm function of social constructivism, and propose a process for fostering norm functions based on the establishment of social norms from the standpoint of cultural evolutionary and social institutional theories. In Chapter 3, we carry out numerical simulation using a resource-sharing problem typical of economic activities involving the production, allocation, and consumption of goods and services from resources in the natural environment, and compare the norm functions fostered through the process proposed in Chapter 2 with a few other predefined norm functions, in terms of society's total value, resource productivity, and the Gini coefficient. In Chapter 4, we assess the simulation results obtained in Chapter 3 in terms of statistical theories on asset distribution in physical economics, and discuss the ideal state of an economic society needed to suppress disparity and inequality, while considering social trends and history of currencies and values. Finally, in Chapter 5, we summarize our conclusions and discuss future issues and prospects that aim for sustainability and fairness.

## 2. Decision-Making Model

### 2.1 Value Function

A major decision-making theory in microeconomics is the rational choice theory. It is based on the principle that individuals choose the rational action that maximizes utility based on methodological individualism. This theory is related to Adam Smith's Invisible Hand, one of the mainstream schools of thought in economics that claims society will attain its optimum state if individuals pursue their self-interests. All these schools of thought will not be discussed here, nevertheless, this theory has been applied to the expected utility theory, the subjective probability theory, the prospect theory, and other theories pertaining to decision-making under conditions of uncertainty and risk, and has been widely used in game theory as an application into sociology and political science, as well as in social choice theory, public choice theory, and comparative institutional analysis.

In the standard rational choice model, during decision-making, an individual assigns a confidence level to beliefs in particular states, allocates cardinal priority criteria to desires for particular results, and maximizes the expected utility of the action. In the decision tree model well-known for decision making, the individual prunes the $s$ ($state$) based on beliefs, and the $o$ ($outcome$) based on desires. The individual utility function ($u(a)$) is expressed in equation (1) below, i.e., for every outcome ($o$), multiply the utility of $o$ ($u(o)$) by the probability of $o$ given $a$ ($action$), ($p(o|a)$), then add these all up (Heath, 2008). The individual chooses the action ($a$) that maximizes the utility function ($u(a)$).

$$u(a) = \sum_o p(o|a) \, u(o)$$

(1)

As an argument against the rational choice theory, some authors claim that it does not take into consideration that an individual sometimes makes altruistic and obligatory actions against his/her self-interests. For example, the experimental game theory pertaining to the collective action problem, has proven that subjects exhibit cooperation and coordination at levels considerably higher than those predicted by the standard model. This is because humans learn socially and follow social norms, including language, customs, and culture, through imitative conformity (Heath, 2008). From the standpoint of economics, the 'opportunity set' of the individual's action is greater than the 'budget set' of the goods, and the Invisible Hand theorem does not always lead society to its optimum state. It can be said, however, that society is established because social norms restrict the 'opportunity set' of the individual's actions (Basu, 2010).

In an attempt to incorporate social norms into rational choice models, the addition of an increase in utility to the kindness of others (Rabin, 1993), or the reduction of utility to the degree of inequality between oneself and others (Fehr & Schmidt, 1999) have been considered. However, modifications that add the benefits of interaction with others to the utility function can not account for a wide range of norms and anonymous cooperative behavior. Furthermore, among the several versions of reciprocity altruism, direct reciprocity can not explain broad sociality without direct relationship. Indirect reciprocity has a primary dilemma (why cooperate?) and a secondary dilemma (why do expensive sanctions?). In order to solve these dilemmas, it is necessary to further assume higher order sanctions, or to link primary and secondary cooperative actions (Henrich & Boyd, 2001; Yamagishi and Takahashi, 1994), however, these dilemmas remain unexplained. Strong reciprocity describes the norms of cooperation by assuming the willingness to do costly altruistic punishment as a result of group selection (Bowles & Gintis, 2011), but this cannot explain other norms such as fairness. Empirical studies of reciprocity have shown that high-order sanctions and heavy punishment are not observed, and that inexpensive sanctions (e.g. break-off of relations, light attention, etc.) are the main consequences that are observed (Kiyonari & Barclay, 2008; Guala, 2012).

Therefore, in incorporating social norms into rational choice models, we would like to adapt two major perspectives: deontology and utilitarianism. In deontology according to Immanuel Kant, humans are expected to follow universal moral rules dictated by reason, good will is an action only based on following one's faith, whereas norms and ethics do not be reduced to utility. On the other hand, in utilitarianism, the social desirability of an action is determined by its' utility, and the goal is to maximize the summation of individual utility ("the greatest happiness of the greatest number"), and norms and ethics are embedded in utility.

Although deontology is based on reason, this reason or rational thinking, from the perspective of social constructivism, is socially constructed along with norms, such as language, customs, and culture. Norm conformity develops through imitative learning during childhood and social learning from the cultural environment, and cannot be isolated from the rational subject. A simple method for incorporating deontological constraints for actions into the rational choice model, in the same way as with utilities for desires, is by handling norms based on reason as instruments (Heath, 2008). In accordance with Savage's trichotomy (states, actions and outcomes), the normative principle connected to the actions are conceived in the same way as the beliefs connected to the states and the outcomes connected to the desires, and normative appropriateness is assigned as weight in considering actions. Therefore, the individual's value function ($v(a)$) can be expressed as the sum of utility ($u(a)$) and normative appropriateness ($n(a)$), as shown in equation (2).

$$v(a) = u(a) + n(a)$$

(2)

From the perspective based on utilitarianism, examples of social norms are community norms that serve as self-enforcing solutions to the commons problem (Aoki, 2001). In this case, the emergence of community norms can be seen by linking the commons game and the social exchange game. The value function ($v(a)$) can be expressed from the utility ($u(a)$) and cooperation cost ($C(a)$) in the commons domain, such as in common water supplies and commonly owned forests, and the utility ($u_s(a)$) and cooperation cost ($C_s(a)$) in the social exchange domain, such as in mutual aid and cooperative, as shown in equation (3). The incentive condition for cooperation can be expressed by equation (4), with the current cooperation cost saving terms on the left-hand side, and the current value conversion (where $\delta$ is the discount factor) of the loss term arising from permanent social ostracization in the future on the right-hand side. When this condition holds true, the common expectation against social ostracization of neglects generates a cooperative community norm.

$$v(a) = u(a) - C(a) + u_s(a) - C_s(a)$$

(3)

$$C(a) + C_s(a) < \delta \cdot u_s(a)$$

(4)

We derived the mathematical equations (2) and (3) for a rational choice model that incorporates social norms from both the standpoints of deontology and utilitarianism. In deontology, norms are shown as intrinsic values by instrumentally treating deontic constraints due to reason. On the other hand, in utilitarianism, the utility of the social exchange that underlies the norm is shown as extrinsic value. Their mathematical expressions, although having different premises, are similar and can therefore be treated equally when incorporating them as information model for social practices. In equation (2), when the preference order for $a$ relative to $n(a)$ is opposite that of the preference order for $a$ relative to $u(a)$, or when $u(a)$ is expressed as an increasing function of $a$, and $n(a)$ is expressed as a decreasing function of $a$, then $n(a)$ becomes a cost and constraint. In equation (3), $-C(a)$ and $-C_s(a)$, to begin with, are costs and constraints relative to utility function $u(a) + u_s(a)$. Since $n(a)$ in equation (2) and $-C(a) - C_s(a)$ in equation (3) can be treated equally, we will use mainly equation (2) in discussions hereinafter. If the normative cost ($n(a)$) is redistributed as tax to society, or $n(a)$ is assumed as cooperating cost ($C(a) + C_s(a)$) in social exchange, deontology can also be regarded as being reduced to social utility.

*2.2 Norm-Fostering Process*

Individuals are not born with inherent social norms but acquire them through norm conformity and norm functions are formed separately for each person. How, then, are norm functions formed? The establishment of social norms can be viewed from two major standpoints; namely, the cultural evolutionary theory and the social institutional theory. The cultural evolutionary theory explains the establishment of norms as part of the dual inheritance system that arises from the coevolution of genetic/biological transmission and cultural/social transmission. The social institutional theory explains the establishment of norms as part of an institutional system arising from a cycling between the individual's propensity and actions, and the group's conditions and symbolisms.

In accordance with the gene-culture dual inheritance theory of Peter J. Richerson and Robert Boyd, Joseph Heath claims that humans learn norms along with language, customs, and culture through a genetically endowed propensity toward conformity and imitative social learning based on those propensities (Richerson & Boyd, 2005), and that these norms cannot be isolated from the human rational and intentional thinking. Moreover, Heath also argues that explicit rules are merely derived from norms based on regulism, which equates norms with explicit rules (e.g. signs), and that arbitrary boundaries can be made for actions to an unlimited extent based on regularism, which equates norms with regularities of actions (e.g. behavior patterns). In accordance with Robert Brandom's interpretation of the origin of norms (Brandom, 1994), norms are enforced by the structure for mutual expectations and reciprocal sanctions based on the norm conformity concept forwarded by Richerson and Boyd (Heath, 2008).

Carsten Herrmann-Pillath and Ivan Boldyrev revised and expanded the institutional model based on Masahiko Aoki's comparative institutional analysis (Aoki, 2001) to present a recursive institutional system wherein: individual propensities trigger actions, the summation of the interaction of individual actions generates a state of equilibrium as the group's shared expectation, a symbol system is then brought about as the summary expression of the state of equilibrium, and the symbols affect the individual propensities. Similar to Brandom's Hegelian solution (Brandom, 2009), the basis for the generation of the group's state of equilibrium from the interaction of individuals lies in the mutual recognition between the subject of self and of others (recognition), wherein: the human mind is composed of the subject (agent) and the object (action), the subject performs actions as objects (performativity), and the objects influence the subjects in a manner that establishes social consistency (continuity) (Herrmann-Pillath & Boldyrev, 2016).

Summarizing the views of Heath, Herrmann-Pillath and Ivan Boldyrev, we can say that social norms are fostered through the cyclic repetition of mutual expectation and recognition between the self and others. We therefore propose the model shown in Figure 1 as a process for fostering norm functions. The individual value function ($v(a)$) shown in equation (2) consists of the individual utility function ($u(a)$) and social norm function ($n(a)$), where the utility function ($u(a)$) is determined individually, whereas the norm function ($n(a)$) is influenced by the mutual expectation and recognition with others. In Figure 1, individual $i$ with the norm function $n_i(a_i)$ sends expectations and recognitions to individual $i-1$ with norm function $n_{i-1}(a_{i-1})$ and individual $i+1$ with norm function $n_{i+1}(a_{i+1})$, etc., and also likewise receives expectations and recognitions from individual $i-1$ and individual $i+1$, etc. In other words, individual $i$, individual $i-1$, individual $i+1$, etc., which are connected in a social relationship network, compare each other's norm functions. And, through a cyclic repetition of mutual expectation and recognition, as shown in equation (5), the norm function $n_i(a)[t]$ is gradually updated into norm function $n_i(a)[t+1]$ at a certain point for each individual or after $t$ number of repetitions, progressively converging into $n_i(a)[\infty]$, wherein norms are fostered as the group's shared expectation, eventually generating the group's state of equilibrium. As such, it will be possible to perform numerical simulation for cardinal and concrete problems based on the norm-fostering process shown in Figure 1.

$$n_i(a)[t] \ \rightarrow \ n_i(a)[t+1] \ \rightarrow \ n_i(a)[\infty]$$

(5)

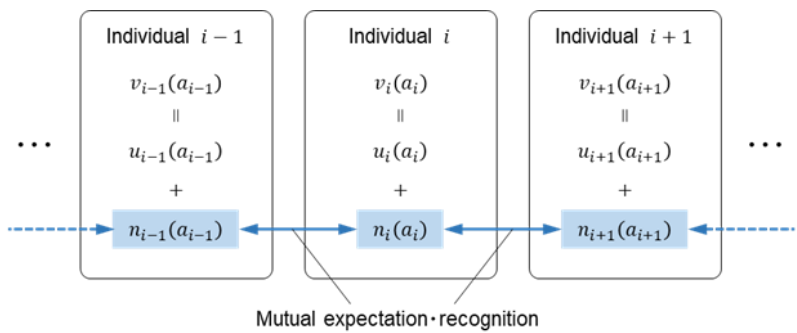

Figure 1. Norm-fostering process based on mutual expectation and recognition

## 3. Resource-Sharing Problem

### 3.1 Problem Setting

Broadly speaking, economic activities pertain to the acquisition of resources from the natural environment, and to the production, allocation, and consumption of goods and services. To determine the effects of social norms on sustainability, fairness, disparity, amongst other social norms, we looked at the resource-sharing problem as a typical example of economic activities. The resource-sharing problem is a typical example pertaining to the competition for shared resources related to disparity and inequality in production activities, allocation of resources and energy, traffic congestion, and the supply chain. It is also a problem that encompasses the tragedy of commons, which deals with the competition between producers for shared resources (the public goods game, that deals with the contribution problem for cooperation cost for public benefits) and the community norm problem based on the linking game for public goods and social exchange (Aoki, 2001). It can therefore be considered as a mathematical model for the cardinal computation of these problems.

The resource-sharing problem is based on the use of shared resources by each individual to gain profits. Since the resource unit price increases when the total amount of used resources increases, a state of equilibrium in the group arises between the maximization of self-value through the use of more resources by each individual, and the increase in the resource unit price of the entire group. This state of equilibrium arises when all individuals try to maximize self-value. The resource-sharing problem can be expressed as a distributed constraint optimization problem as shown in equation (6), where $N$ is total number of persons in the group, $x_i$ is the resource used by individual $i$ ($i = 1, 2, \cdots, N$), $f_i(x_i)$ is the value function, and $g$ is the total cost function determined by the total used resources (e.g. Boyd, Parikh,

Chu, Peleato & Eckstein, 2011).

$$\arg\max_{\{x_i\}} \sum_{i=1}^{N} f_i(x_i) - g\left(\sum_{i=1}^{N} x_i\right)$$

(6)

To set a concrete problem, $x_i$ is expressed as the resource used by individual and $Z$ as the total resources used by the group (equation (7)), and unit price for resource is expressed as $p$ (equation (8)). The resource unit price $p$ is the result of adding the constant $c$ to the product of the coefficient $b$ and total resource $Z$ raised to the power of $r$. Exponent $r$ is $r > 1$, and expresses the effect of increasing costs. As shown in equation (9), the profit, i.e., the utility function ($u_i$), of individual $i$ is the result of subtracting the product of used resources ($x_i$) and resource unit price ($p$) from the product of resource ($x_i$) raised to the power of $s$ and action ($a_i$) of individual $i$. Action ($a_i$) of individual $i$ can be interpreted as the production capacity relative to the resources, as the potential for handling the resources, or as the effort needed to obtain profit from the resources. Exponent $s$ is $s < 1$, and expresses the effect of diminishing returns. As shown in equation (10), the combination of equations (7) to (9) corresponds to equation (6), which represents the original distributed optimization problem.

$$z = \sum_{i=1}^{N} x_i$$

(7)

$$p = b \cdot z^r + c$$

(8)

$$u_i = a_i \cdot x_i^s - p \cdot x_i$$

(9)

$$\arg\max_{\{x_i\}} \sum_{i=1}^{N} u_i = \arg\max_{\{x_i\}} \sum_{i=1}^{N} a_i \cdot x_i^s - \sum_{i=1}^{N} x_i \cdot \left\{ b \cdot \left(\sum_{i=1}^{N} x_i\right)^r + c \right\}$$

(10)

Thus far, we have shown equations using only utility function ($u_i$). To incorporate social norms, we will replace the utility function ($u_i$) shown in equation (9) with the value function incorporating norm function shown in equation (2). To compare the effect of norm, we set three value functions ($v_{1_i}, v_{2_i}, v_{3_i}$) as shown in equations (11) to (13).

For the value function $v_{1_i}$ in equation (11), we set the norm function multiplied with the norm coefficient $n_{1_i}$, which differs for each individual ($i$) relative to used resource ($x_i$). This norm coefficient ($n_{1_i}$) is fostered through mutual expectation and recognition with others, as shown in Figure 1 and equation (5). As will be demonstrated in the results of simulation in Section 3.2, the last term of equation (11) can be considered as a progressive tax-like cost (however, it should be noted that the result is a progressive tax-like, not an ex-post cost redistribution by tax collection, but an ongoing cost distribution by interaction between individuals).

For $v_{2_i}$ in equation (12), we set the norm function multiplied with the norm coefficient $n_2$, which is a constant ratio relative to used resources ($x_i$). The last term of equation (12) is a proportional tax-like (consumption tax) cost.

For $v_{3_i}$ in equation (13), we set the norm coefficient $n_3$, which is a fixed value regardless of used resources ($x_i$). The last term of equation (13) is a fixed tax-like cost.

Using criteria based on the equity theory (Adams, 1963) as the criteria for mutual comparison, as show in equation (14), through the norm-fostering process, the value/action ratio ($v_{1_i}[t]/a_i$) for individual ($i$) is mutually compared with $m$ number of other persons connected in a social relationship network (shown in Figure 3 below), and the norm ($n_{1_i}[t]$) is gradually updated into ($n_{1_i}[t + 1]$) in Δ steps

The sum of the normative cost for the group (the so-called total tax revenue) is expressed in equations (15) to (17) respectively for value functions $v_{1_i}, v_{2_i}, and\ v_{3_i}$ of equations (11) to (13).

$$v_{1_i} = u_i - n_{1_i} \cdot x_i = a_i \cdot x_i^s - p \cdot x_i - n_{1_i} \cdot x_i$$

(11)

$$v_{2_i} = u_i - n_2 \cdot x_i = a_i \cdot x_i^s - p \cdot x_i - n_2 \cdot x_i \tag{12}$$

$$v_{3_i} = u_i - n_3 = a_i \cdot x_i^s - p \cdot x_i - n_3 \tag{13}$$

$$\frac{v_{1_i}[t]}{a_i} \gtreqless \frac{1}{m} \cdot \sum_{j \neq i}^{m} \frac{v_{1_j}[t]}{a_j} \Rightarrow n_{1_i}[t+1] = n_{1_i}[t] \pm \Delta \tag{14}$$

$$w_1 = \sum_{i=1}^{N} n_{1_i} \cdot x_i \tag{15}$$

$$w_2 = \sum_{i=1}^{N} n_2 \cdot x_i \tag{16}$$

$$w_3 = n_3 \cdot N \tag{17}$$

Although the resource-sharing problem is a distributed optimization problem for a group, it is a rational choice problem for maximizing value functions ($v_{1_i}, v_{2_i}, v_{3_i}$) for each individual ($i$), and a problem for finding the optimum value of used resources ($x_i$) relative to the individual ($i$) action ($a_i$) and resource unit price ($p$). Optimum values of used resources ($x_i$) for each of the value functions ($v_{1_i}, v_{2_i}, v_{3_i}$) in equations (11) to (13) are shown in equations (18) to (20), respectively. However, since resource unit price ($p$) changes depending on the resources used by others, there is a need to gradually update the optimum values until the group reaches a state of equilibrium.

As shown in Figure 2, in the computational flow for value function $v_{1_i}$ of equation (11), firstly, the total resource ($z[t+1]$) is computed from used resource ($x_i$ ($i = 1, 2, \cdots, N$)) for a certain point of time or for $t$ number of repetitions, then the resource unit price ($p[t+1]$) is computed from $z[t]$ and $z[t+1]$, the norm ($n_{1_i}[t]$) is subsequently updated to ($n_{1_i}[t+1]$) in $\Delta$ steps based on individual ($i$) value/action ratio ($v_{1_i}[t]/a_i$) and on mutual comparison with others following equation (14). Following this, the optimum value of used resource ($x_i$) for point $t+1$ is computed following equation (18), recursively repeating the entire process until the optimum value converges. Herein, finding resource unit price ($p$) from not only $z[t+1]$, but the sum of $z[t]$ and $z[t+1]$, is done for computational reasons; namely, to prevent sudden fluctuations in resource unit price and hasten the convergence. Moreover, the computational flows for value functions $v_{2_i}$ and $v_{3_i}$ of equations (12) and (13) are similar to the computational flow shown in Figure 2, except for the absence of update of norm coefficients and the computation of optimum value following equations (19) and (20), respectively.

$$x_i = \left(\frac{s \cdot a_i}{p + n_{1_i}}\right)^{\frac{1}{1-s}} \tag{18}$$

$$x_i = \left(\frac{s \cdot a_i}{p + n_2}\right)^{\frac{1}{1-s}} \tag{19}$$

$$x_i = \left(\frac{s \cdot a_i}{p}\right)^{\frac{1}{1-s}} \tag{20}$$

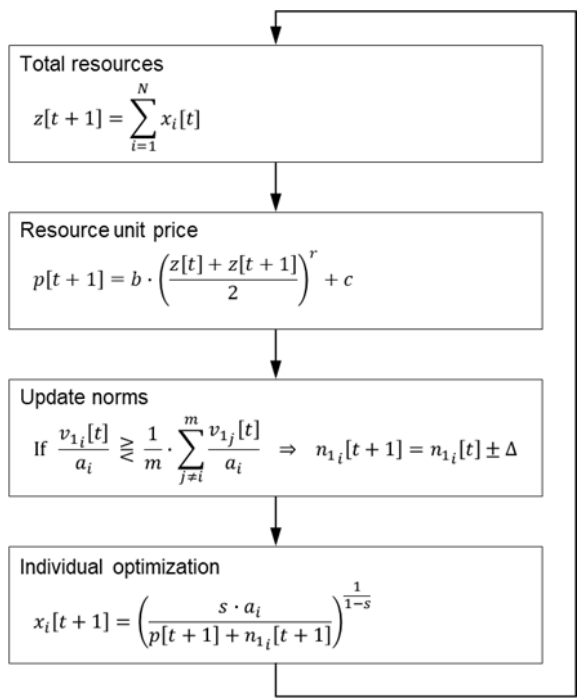

Figure 2. Computational flow of norm-fostering process for resource-sharing problem

*3.2 Numerical Simulation*

We conducted actual numerical simulation by setting the total number of persons in the group to 100 ($N = 100$), and set the social relationship network for individual ($i$ ($i = 1, 2, \cdots, N$)) assuming a scale-free network topology, as shown in Figure 3. A scale-free network topology is commonly seen in Internet and literature citation relationships, as well as in social relationships. Figure 3 is an example of a scale-free network with 100 nodes and 2 degrees generated using the famous Barabási-Albert model.

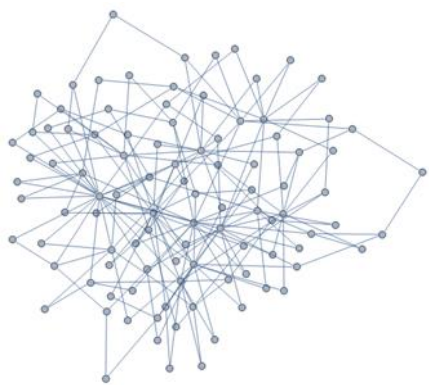

Figure 3. Social relations network (Barabási-Albert model, 100 nodes, 2 degrees)

The distribution of actions ($a_i$ ($i = 1, 2, \cdots, N$)) of individual ($i$) is set as two types of distribution, namely, the normal distribution and power-law distribution, as shown in Figure 4. For the normal distribution, which is widely observed for academic grades, body height, and production capability, the average value for action ($a_i$) is set to $\bar{a} = \mu = 0.5$, standard deviation to $\sigma = 0.1$, and histogram bin to $\Delta a = 0.025$. The power-law distribution is a long-tail distribution widely observed for populations, incomes, and assets. The probability density function for power-law distribution is set to $1/4 \cdot a^{-k}$, and $k = 2$, since the scaling exponent for income or asset distribution is empirically 2. Additionally, since the average value has no meaning in the power-law distribution, the median value is set to $\tilde{a} = 0.5$ in accordance with the average value ($\bar{a}$) in the normal distribution, and the histogram bin to $\Delta a = 0.025$ similar to that of the normal distribution.

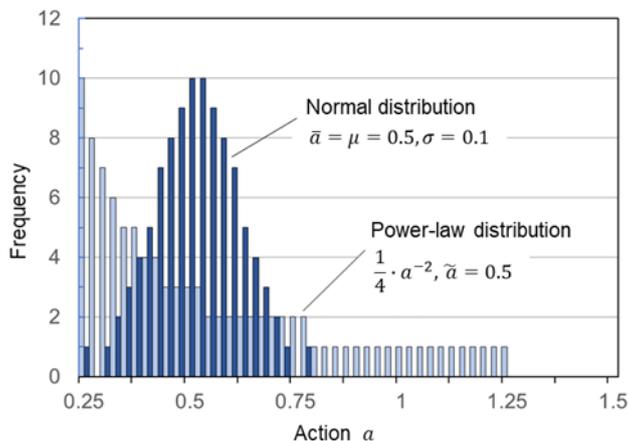

Figure 4. Histograms of normal distribution and power-law distribution of actions

The calculation parameter set points are set to: equation (8) i.e. $p = b \cdot z^r + c$ with coefficient $b = 0.001$, constant term $c = 1.0$, exponent $r = 2.0$, equation (11) to (13) exponent $s = 0.5$, and equation (14) step $\Delta = 0.05$. And, all computational conditions are prepared by setting the initial value to $z[0] = 1.0$ for $t = 0$ of total resource $Z$, enabling simulation following the computational flow in Figure 2.

Figure 5 shows the results of simulation using value function $v_{1_i}$ (progressive tax-like, value/action ratio) of equation (11), $v_{2_i}$ (proportional tax-like) of equation (12), and $v_{3_i}$ (fixed tax-like) of equation (13), for the normal distribution and power-law distribution of individual ($i$) action ($a_i$). The three graphs (a) to (c) on the left side of Figure 5 follow a normal distribution, while the three graphs (d) to (f) on the right side follow a power-law distribution, with the upper two graphs (a) and (d) corresponding to value function $v_{1_i}$, the middle two graphs (b) and (e) to $v_{2_i}$, and the lower two graphs (c) and (f) to $v_{3_i}$. The vertical axis of each graph indicates value function ($v_{1_i}, v_{2_i}, v_{3_i}$ ($i = 1, 2, \cdots, N$)), and the horizontal axis indicates repetition number ($t$). To be able to compare between the same type of distribution, for the normal distribution, we adjusted the values of $n_2$ and $n_3$ in accordance with the computation results for $w_1$ so that the sum of normative cost (total tax revenue) shown in equations (15) to (17) would be equal, i.e., $w_1 = w_2 = w_3 \simeq 1.85$. Likewise, for the power-law distribution, we adjusted the values so that $w_1 = w_2 = w_3 \simeq 1.62$. All the graphs showed that the value for value functions ($v_{1_i}, v_{2_i}, v_{3_i}$) converged as the number of repetitions ($t$) for recursive computation increased.

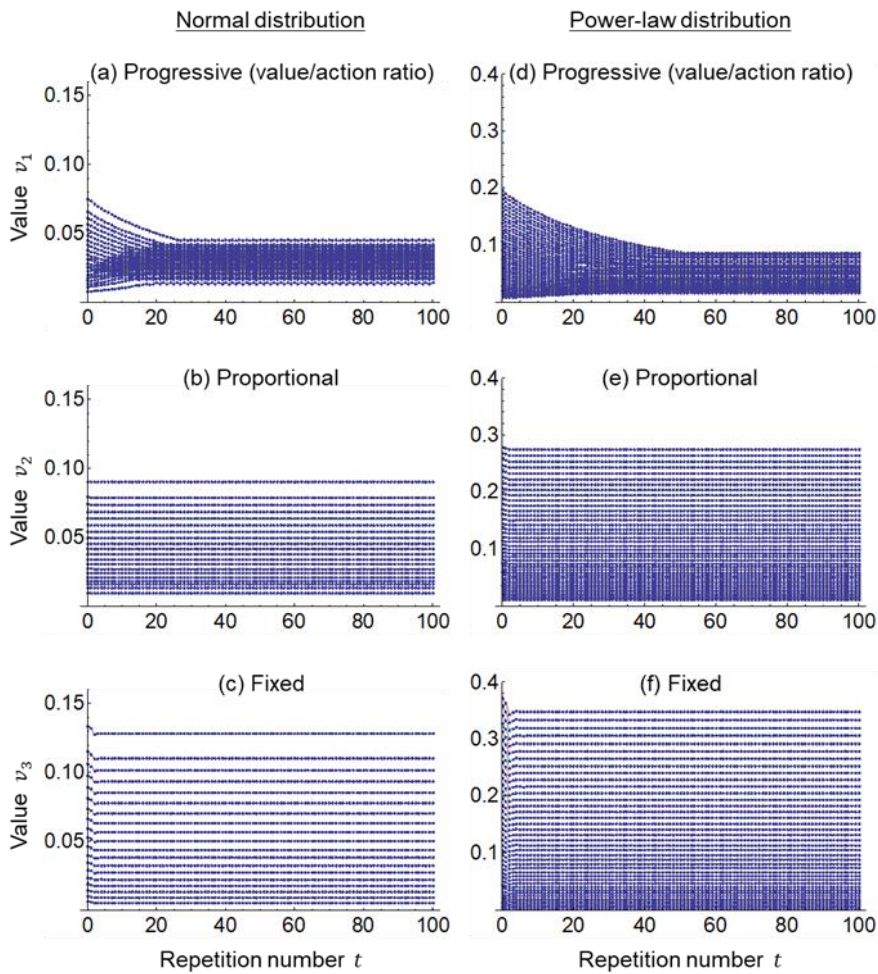

Figure 5. Simulation results of value functions based on recursive computational flow

Figure 6 shows histograms for distribution of values for value functions $(v_{1_i}, v_{2_i}, v_{3_i})$ for $t = 100$ repetitions, where recursive computations have fully converged. The four graphs (a) to (d) on the left side of Figure 6 follow a normal distribution, while the four graphs (e) to (h) on the right side follow a power-law distribution, with the first (uppermost) two graphs (a) and (e) showing the original distribution of action $(a_i)$, the second two graphs (b) and (f) show the distribution for value function $v_{1_i}$ (progressive tax-like), the third two graphs (c) and (g) for $v_{2_i}$ (proportional tax-like), and the fourth (lowermost) two graphs (d) and (h) for $v_{3_i}$ (fixed tax-like).

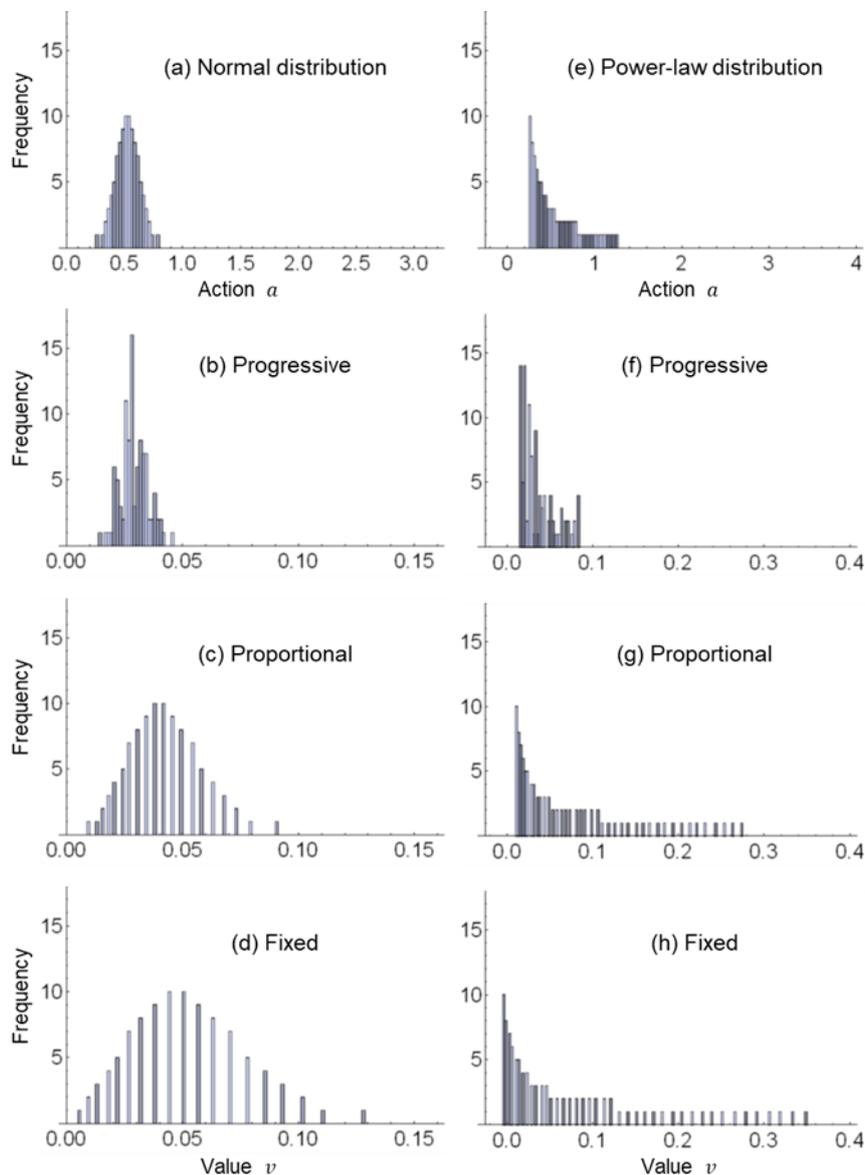

Figure 6. Histograms of individual action distributions and value function distributions

From Figures 5 and 6, in contrast to the distribution of individual ($i$) action ($a_i$), for both normal and power-law distributions, the distribution of value ($v$) widened as norms shifted from progressive tax-like norm ($n_{1_i}$), to proportional tax-like norm ($n_2$), and to fixed tax-like norm ($n_3$), i.e., the disparity among individuals widened. Comparing normal distribution and power-law distribution showed that the power-law distribution had a wider spread compared to the normal distribution.

Plotting the value ($v_{1_i}, v_{2_i}, v_{3_i}$) against action ($a_i$) showed that value $v_{1_i}$ was proportional to action ($a_i$) for the progressive tax-like norm ($n_{1_i}$), whereas values $v_{2_i}$ and $v_{3_i}$ exhibited the square function of action ($a_i$) for the proportional tax-like norm ($n_2$) and fixed tax-like norm ($n_3$), indicating an expanding disparity for these two norms as shown in Figure 7.

Values ($v_{2_i}, v_{3_i}$) exhibit square functions for norm $n_2$ and norm $n_3$ because the exponent ($1/(1-s)$) ($s = 0.5$) in equations (19) and (20) is a square. This means that when the utility function diminishes, the disparity tends to widen.

But, values $v_{1_i}$ for norm $n_{1_i}$ did not exhibit a square function but a linear function because the normative cost based on the fairness criteria increased proportionally relative to the action ($a_i$), which in turn suppressed disparity. However, the normative costs for norm $n_{1_i}$ and norm $n_2$ were almost equal; indicating that disparity might also be suppressed in norm $n_2$, as shown in Figure 8.

However, that in the upper end of the distribution of action ($a_i$), norm $n_2$ had a significantly smaller norm coefficient when compared to norm $n_{1_i}$, which was the opposite for the lower end of the distribution. Because of this, as is evident in equations (11) and (12) and equations (18) and (19), values $v_{2_i}$ for norm $n_2$ enlarged to almost the square of action ($a_i$) in the upper end of the distribution of action ($a_i$) compared to norm $n_{1_i}$, while, conversely, normative cost enlarged to almost the square of action ($a_i$) in the lower end of the distribution, as shown in Figure 9. Disparity, therefore, was wider for the proportional tax-like norm ($n_2$) than for the progressive tax-like norm ($n_{1_i}$), as shown in Figure 7.

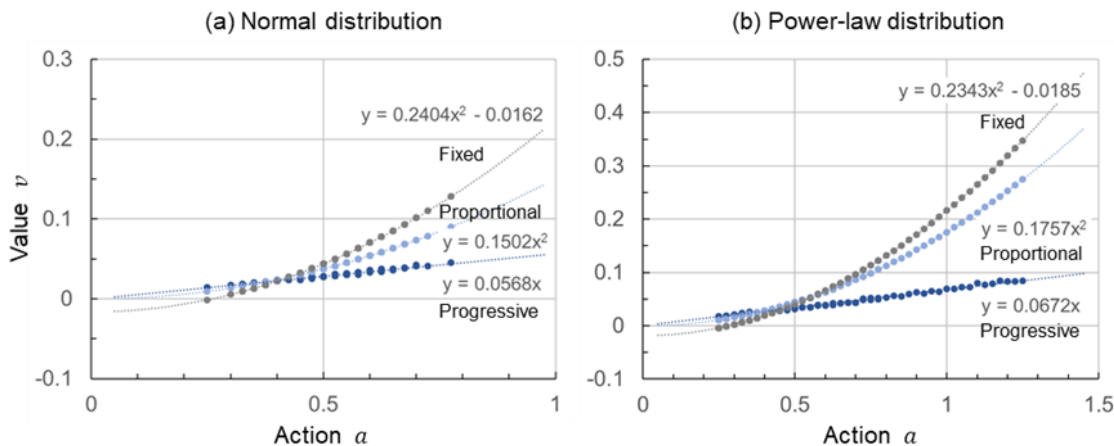

Figure 7. Comparison of dependence of value functions on actions of individuals

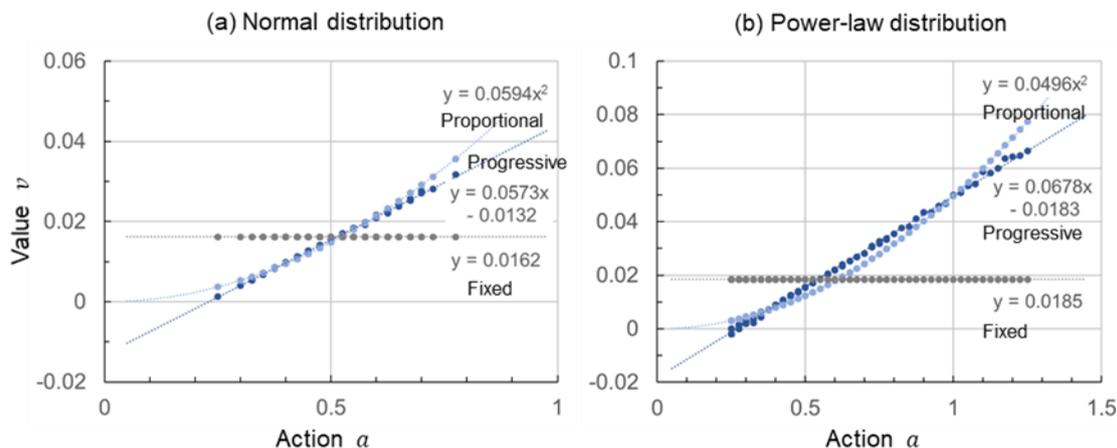

Figure 8. Comparison of dependence of normative costs on actions of individuals

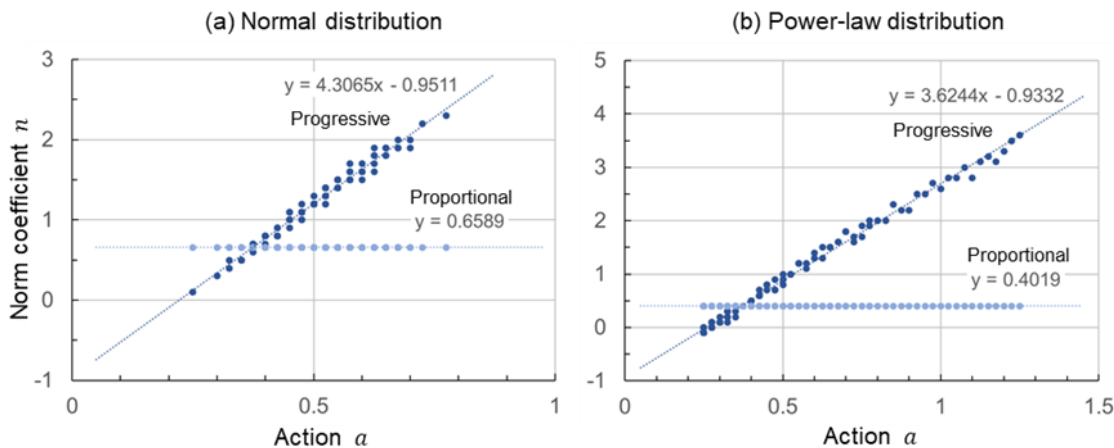

Figure 9. Comparison of dependence of norm coefficients on actions of individuals

Figure 10 (a) shows results of computation of total value (sum of value functions $v_{1_i}, v_{2_i}, v_{3_i}$) and Figure 10 (b) shows total resources ($z[100]$) for $t = 100$ repetitions for a population of 100 persons. Both the total value and total resources became larger as norms shifted from progressive tax-like norm ($n_{1_i}$), to proportional tax-like norm ($n_2$), and to fixed tax-like norm ($n_3$), and total value and total resources in the power-law distribution was larger than in the normal distribution. These tendencies could be attributed to having individuals in the upper end of the distribution of action ($a_i$) using more resources and gaining more value, for the fixed tax-like norm $n_3$ and proportional tax-like norm $n_2$ compared to the progressive tax-like norm $n_{1_i}$, as well as for the power-law distribution compared to the normal distribution. In other words, the fixed tax-like norm $n_3$ and proportional tax-like norm $n_2$ tended to expand disparity, particularly in a power-law distribution.

Figure 11 (a) shows the resource productivity (total value/total resource ratio) computed from the total value and total resources shown in Figure 10 and Figure 11 (b) shows the Gini coefficients computed from distribution of value ($v$). Progressive tax-like norm ($n_{1_i}$) had the highest resource productivity compared to fixed tax-line norm ($n_3$) and proportional tax-like norm ($n_2$). Compared to the other two norms, progressive tax-like norm ($n_{1_i}$) had the lowest Gini coefficient, which is the same as the Gini coefficient of the original distribution of action ($a_i$) (dotted line in the figure). This meant that progressive tax-like norm ($n_{1_i}$) had the highest sustainability in terms of the efficient use of resources, and had the highest fairness and suppression of disparity in terms of maintaining and preventing the increase of the Gini coefficient. Moreover, although this section only presented results of simulation for a single set of conditions for social relationship network, individual action distribution, and calculation parameters, it should be noted that the tendencies for the results given in this section would remain the same even if the conditions were changed.

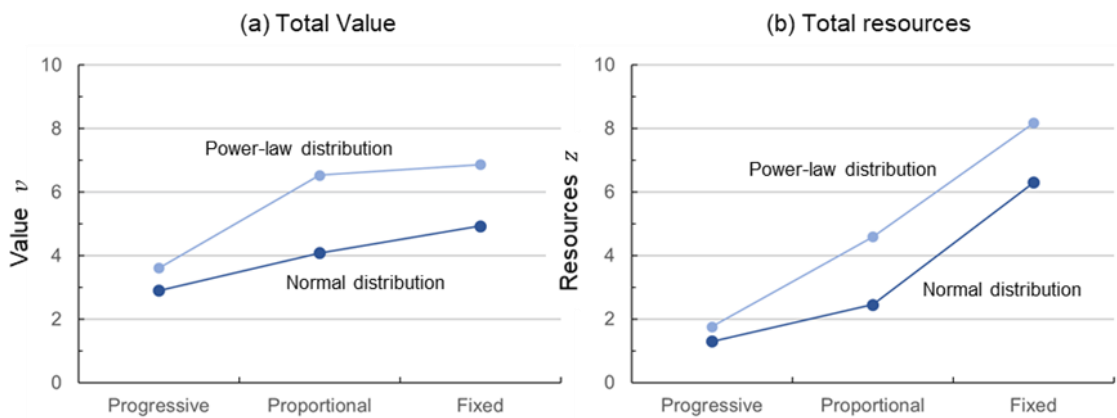

Figure 10. Comparison of total values and total resources for action distributions and norms

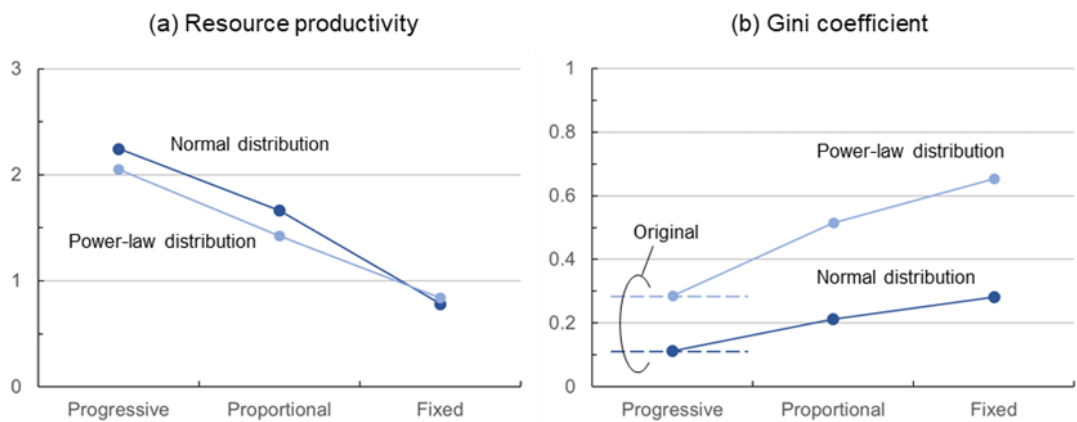

Figure 11. Comparison of resource productivities and Gini coefficients

## 4. Discussions

*4.1 Review of Results*

Results of numerical simulation using the resource-sharing problem showed that the progressive tax-like norm function fostered through the mutual comparison of value/action ratio based on the equity theory, had higher resource productivity and lower Gini coefficient compared to proportional tax-like or fixed tax-like norm functions. Progressive tax-like norm function is therefore the most preferable norm function in terms of sustainability and fairness. Taking into consideration the negative correlation between the Gini coefficient and the human development index developed by Amartya Sen and Mahbub ul Haq (Holden, Linnerud & Banister, 2014), the progressive tax-like norm function is also preferable in terms of the individual's capabilities. In the proportional tax-like and fixed tax-like norm functions, the distribution of value widens proportionally to a square of the distribution of individual actions, with a tendency to widen the disparity between individuals since individuals in the upper end of the distribution of actions use more resources and gain more value than those in the lower end. The consumption of total resources would therefore also increase along with the increase in the total value of the group. However, in order to realize sustainable, fair, and inclusive societies, for instance by reducing income disparity and inequalities, ensuring access to energy, protection of the environment, and other Sustainable Development Goals (SDGs) adopted by the United Nations (United Nations, 2018), not only should utility in terms of value in society be taken into consideration, but also fostering progressive tax-like norms. The progressive tax-like norms discussed herein is based on individual values fostered in the ongoing interaction between individuals, and that eventually bring a progressive tax result at a group's state of equilibrium and not ex-post social security.

As regards the effect of the distribution of actions, results of simulation showed that the normal distribution had only a slightly higher resource productivity and a significantly lower Gini coefficient than power-law distribution. Because the power-law distribution has wider spread than the normal distribution, there is a tendency for those in the upper end of the distribution of actions to also be in the upper end of the distribution of values. Even with the normal distribution, the disparity in the distribution of values is wider for proportional tax-like and fixed tax-like norms than for the progressive tax-like norms. In cases when the value obtained in the current period becomes an asset and affects the action in the next period, it can be easily inferred that the normal distribution will also eventually approach a widely spread distribution similar to the power-law distribution. The random-sharing model, where assets are randomly shared between two parties, exhibits an exponential distribution (A. S. Chakrabarti & B. K. Chakrabarti, 2010), whereas the model for performing exchange and saving of assets exhibits a shift from a distribution similar to a normal distribution to a distribution similar to a power-law distribution as the ratio of the savings to assets decreases (change in gamma distribution parameter) (Angle, 2006). Consequently, as long as assets are exchanged, in the initial period of the resource-sharing problem, the distribution of actions approaches the power-law distribution, if the value obtained in the current period becomes an asset, the disparity in value and assets in the subsequent periods will recursively expand for proportional tax-like and fixed tax-like norms. These relationships are similar to the r>g (return on capital > economic growth) configuration shown by Thomas Piketty in regard to capitalization rate and economic growth rate (Piketty, 2014).

Although it is possible to maintain and prevent the increase of disparity of the original action distribution, in the progressive tax-like norm based on the fairness criteria (value/action ratio), it is not possible to modify the original action distribution itself. To modify the original distribution despite having an exchange of assets, would necessitate a norm that is more rigid than the progressive tax-like norm. Would it be possible to foster such a norm? In developmental psychology, humans are said to possess moral norms pertaining to fairness and an ability to differentiate between right and wrong from infancy (Surian, Ueno, Itakura & Meristo, 2018). These are norms such as physical safety and health instincts arise developed to facilitate survival. But, it is unlikely for humans to have instinctive norms more rigid than fairness. In cultural psychology, humans are said to acquire a value system through their families or schools at an early age, i.e. by the attainment of 10 to 12 years of age and it is fairly difficult to change the mental development acquired at this stage (G. Hofstede, G. J. Hofstede & Minkov, 2010). Therefore, humans are not expected to develop norms more rigid than fairness at home or in schools.

From the perspective of human history, the birth of Homo sapiens came about around 200 000 years ago after evolving through the hunting-gathering society. The agricultural society then begun around 8500 years BC after the neolithic age. The exchange of goods was the main practice, and norms on fairness must have already been sufficiently developed in the beginning of the agricultural society. The use of money as a replacement to bartering; namely, the weighing currency, came about as a measure of value around 3300 BC, and the precious metal currency arose as a means to save value around the 5th century BC. Capitalization begun around the end of the 16th century, and asset management around the beginning of the 18th century. Therefore, only a limited proportion (0.2%) of human history has elapsed since the onset of capitalization and asset management. Norms more rigid than fairness were not even recorded during the axial age (around 800 BC to 200 BC), a period known to have produced many philosophers and thinkers. Humans beings therefore, still do not possess the norms needed to properly adapt to capitalization and asset management.

Therefore, we believe that it is imperative, to prevent deviations from occurring in the original distribution of actions, to generate value through norms based on fairness, to increase sustainability, and to suppress disparity. In other words, there is a need to eliminate the conversion of generated values into assets and the exchange of assets, and also to aim for a society where economic activities are carried out based on progressive tax-like norms. A globalized economic society, however, may not be receptive to adapting progressive tax-like norms. Despite Kaushik Basu having explained the importance of international policy coordination aimed to mitigate poverty and inequality, he also spoke of the challenges and the intentions of achieving these aims (Basu, 2010). Currently, there are tax havens that allow the rich to avoid taxes, some countries grant tax relief to attract and nurture companies, and taxation measures centered on consumption tax (proportional tax) are currently being implemented. This illustrates that international coordination on tax policies is still something to be envisaged in the distant future.

Therefore, what should we do? One hope we have is in the transformation from global to local and from this into a regional and community economy (Hiroi, 2009). This is related to patronizing regional cuisine, promoting local production for local consumption, use of renewable energy, focus on local ties and culture, and other advocacies forwarded by Helena Norberg-Hodge, Junko Edahiro, and others (Norberg-Hodge, 2016; Edahiro, 2018). We would like to rethink the three patterns in economic society forwarded by Karl Polanyi (reciprocity, redistribution, and exchange) (Polanyi, 1977) as: reciprocal coordination and sharing through mutual expectation and recognition, redistribution and resetting of assets through progressive tax-like norms, and equal exchange and stockless economic cycling, with a focus on the region and community as the target for social practice, towards the establishment of an ideal, self-reliant local society. Furthermore, by incorporating progressive tax-like norms into the design of regional currency, including the depreciation of money put forward by Silvio Gesell, we believe that we will be able to guarantee sustainability and fairness at least within a region or community.

*4.2 Effects*

Conventionally, decision-making theories have focused only on utility functions, e.g. the game theory measures utility based on the payoff matrix or the logic tree, whereas welfare economics has made logical definitions of institutionalized mechanisms based on axiomatic approaches. In this paper, we introduced a value function that adds the norm function in addition to the utility function, conducted a numerical simulation of a mathematically modeled resource-sharing problem to compute resource productivity and Gini coefficients and compared concrete modalities for social norms enabling us to arrive at suggestions regarding sustainability and fairness. The approach presented in this paper has the potential to contribute to social practices aimed at realizing a society based on social norms, through mutual complementation with decision-making theories, experimental economic theories, welfare economic theories, and other theories.

Our approach, which incorporates both aspects of utility and norm to the resource-sharing problem, can also be applied to real-world problems where there is competition for resources, such as in energy demand and supply, traffic congestion, and the supply chain. A similar approach can also be applied to other problems and distributed optimization problems, such as to the knapsack problem for selecting a variety of commodities, the multi-objective optimization problem for operating passenger buses, and the travelling salesman problem for delivering local services. It therefore may be useful in discussing sustainability, fairness, disparity, and other issues.

For the future society that Japan aspires to realize, the Japanese government has proposed the vision for Society 5.0 which is defined as a "human-centered society that balances economic advancement with the resolution of social problems by a system that highly integrates cyberspace and physical space" (Cabinet Office, Government of Japan, 2014) and succeeds the hunting society (Society 1.0), agricultural society (Society 3.0), industrial society (Society 3.0), and information society (Society 4.0). Before the information society, data analysis, explanation, and prediction of phenomena were performed according to physical paradigms. However, a new paradigm that consists of the diagnosis and prognosis of social phenomena in real time according to clinical medical models and the clinical intervention of social systems by IT systems is proposed to contribute to the realization of Society 5.0 (Deguchi, Otsuka, Kudo & Kato, 2018a).

Society 5.0 aims to solve social problems such as redistribution of wealth and remediation of regional disparities, as well as to enable an active and enjoyable life for everyone. Similarly, the SDGs adopted by the United Nations aim to eradicate poverty and hunger, reduce income disparity and inequality, provide affordable energy and preserve the environment, provide inclusive employment and institutions, amongst other actions to realize a sustainable, diverse, and inclusive society for all (United Nations, 2018). Achieving these goals requires a consideration of social norms, morals, and ethics, such as impartiality, fairness, virtue, and justice, as well as the implementation of normativity and ethics, as opposed to only utility in terms of convenience and efficiency of products and services, as regards IT systems.

Therefore a system is proposed that fuses social systems and IT systems, and in which IT systems perform diagnosis and prognosis of social systems and carry out real-time normative and ethical interventions to social systems based on the identified diagnosis and prognosis, as shown in Figure 12 (Deguchi et al., 2018a; Deguchi, Kato, Kudo, Karasawa & Saigo, 2018b; Karasawa, Kato, Kudo, Yamaguchi, Otsuka & Deguchi, 2018). In this fused system, IT systems carry out normative intervention in different layers; namely, the micro-level individual decision-making and actions, the macro-level interactions between individuals, and the meta-level social institutions. Although the norm-fostering process based on mutual expectation and recognition proposed in this paper mainly relate to the inter-individual interaction layer, individuals are comprised of the subject, the object (action), and a mutual recognition of others. Moreover, institutions are created from recursive cycles of individual actions and group states of equilibrium. The norm-fostering process therefore also relates to individual and institutional layers.

Implementing the norm-fostering process proposed in this paper for IT systems, in the inter-individual interaction layer, will entail the mutual exchange of information on value/action ratios and the mutual transmission of votes and appraisals through IT interfaces, for the individual layer, it will involve nudging and persuasion using behavioral science to promote normative actions and for the institutional layer, it will involve the provision of information pertaining to predicted equilibrium states and scenarios to facilitate the generation of shared expectations. We plan to carry out concrete trials regarding the implementation of IT systems and the modalities of norms explained in this paper through social verification experiments on self-sufficiency of renewable energy and supply chains for local production for local consumption, towards the establishment of self-reliant regional societies that are sustainable, fair, and inclusive.

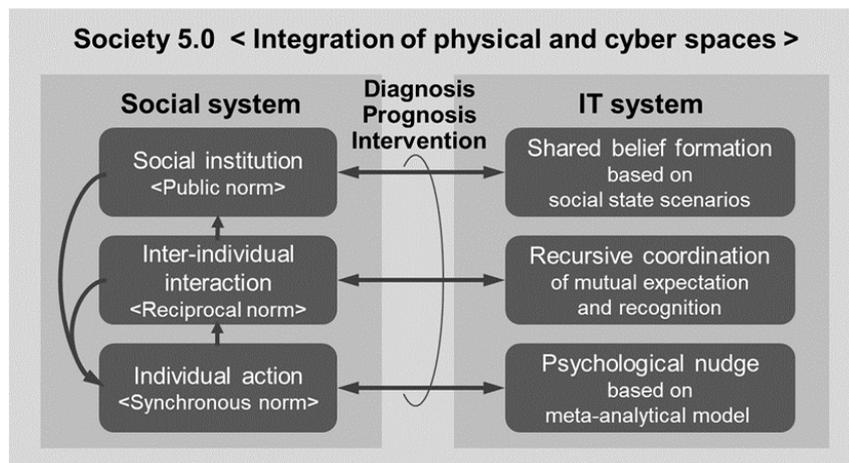

Figure 12. Normative intervention of social systems by IT systems in Society 5.0

## 5. Summary

*5.1 Conclusions*

In an attempt to move towards the realization of a sustainable and fair society, we introduced a decision-making model based on value functions that included utility function and norm function, and proposed a norm-fostering process that recursively updates norm function through mutual expectation and recognition between the self and others. Moreover, we conducted numerical simulation using the resource-sharing problem and showed that the progressive tax-like norm, which is based on the value/action ratio, is preferable over proportional tax-like and fixed tax-like norms in terms of resource productivity (sustainability) and Gini coefficient (fairness).

1) We incorporated the social norm function in addition to the individual utility function into the decision-making model in terms of normative appropriateness and moral cost from the standpoint of deontology, and in terms of social exchange cost and social preferences from the standpoint of utilitarianism. Furthermore, we proposed a norm-fostering process that recursively and gradually updates norm functions while repeating cycles of mutual expectation and recognition. This proposition was done in terms of, social learning and mutual expectation based on conformity from the standpoint of the cultural evolutionary theory, and mutual recognition that serves as basis for the group's shared expectation from the standpoint of social institutional theory.

2) We looked at the resource-sharing problem as a typical example pertaining to the competition for shared resources related to production activities, allocation of resources and energy, traffic congestion, and the supply chain. After estimating the increasing cost function for resource unit price and the diminishing return function for utility, we defined three norm functions for comparative purposes (progressive tax-like, proportional tax-like, and fixed tax-like). In the computational flow for progressive tax-like norm function, the norm function was recursively updated through mutual comparison with others using value/action ratio based on the equity theory as criteria.

3) Results of numerical simulation of the normal distribution and power-law distribution of actions that assumed a scale-free network for a group of 100 persons, showed that for both types of distribution of actions, value is proportional to action in the progressive tax-like norm, exhibits the square function of actions for both the proportional tax-like and fixed tax-like norms, leading to increase in disparity. As regards the group's total value and total resources, both value and resources were larger in the power-law distribution than in the normal distribution, and in the proportional tax-like and fixed tax-like norms than in the progressive tax-like norm; individuals in the upper end of the distribution of actions used more resources and gained more value than those in the lower end of the distribution for the proportional tax-like and fixed tax-like norms. Looking at these results from another perspective, the progressive tax-like norm resulted in the highest resource productivity and the lowest Gini coefficient, meaning that it exhibited the highest sustainability and fairness compared to the other norms.

*5.2 Future Prospects*

We continue to aspire for a global coordination of international policies, as we move on to conduct trials of social practices in local regions and communities, to foster a norm that not only seeks utility in terms of profits or benefits as a social value, but also has excellent sustainability and fairness. Going forward, we intend to take the first step towards achieving sustainability and fairness in a region and a community, rather than as a nation or internationally, by aiming for coordination and sharing based on reciprocal and mutual expectation and recognition, redistribution and resetting of assets through norms, and equal exchange and stockless economic cycling, as well as in the design of institutions and regional currency.

The design and fostering process for norm functions reported in this paper point toa clinical medical prescription and a practical method for intervention of social systems. One of the issues that needs to be addressed subsequently is to separately diagnose utility functions and normative functions from behavioral data of individuals and groups in the real world, and to apply the diagnoses to prescriptions. In addition, research ought to determine the effect free-riders and offenders have on sustainability and fairness. This refers to when the intervention based on the prescription is not effective. Research also ought to investigate the homeostasis in the cycling and maintenance of social groups, to determine the effects of regional and community culture and reevaluate the design of norms and methods of intervention that are suitable to value systems and customs. We endeavor to contribute to the realization of a better society through these activities.